\documentclass[10pt]{cai}
\usepackage{tabularx,multirow,array}
\usepackage[htt]{hyphenat}
\usepackage{seqsplit}
\usepackage{graphicx}
\usepackage[table,dvipsnames]{xcolor}
\usepackage{amssymb}
\usepackage{pifont}
\begin{document}

\def\conferenceyear{2026}
\begin{center}

\title{Adaptive Collaboration of Arena-Based Argumentative LLMs for Explainable and Contestable Legal Reasoning}
\maketitle

\thispagestyle{empty}

\begin{tabular}{cc}
{\small Hoang-Loc Cao\upstairs{\affilone*\S}, Phuc Ho\upstairs{\affilone\S}, Truong Thanh Hung Nguyen\upstairs{\affiltwo},}\\{\small Phuc Truong Loc Nguyen\upstairs{\affilthree},Dinh Thien Loc Nguyen\upstairs{\affilone}, Hung Cao\upstairs{\affiltwo}}
\\[0.25ex]
{\small \upstairs{\affilone}Ho Chi Minh University of Science, Vietnam} \\
{\small \upstairs{\affiltwo}University of New Brunswick, Canada} \\
{\small \upstairs{\affilthree}Friedrich-Alexander-Universität Erlangen-Nürnberg, Germany} \\
\end{tabular}
  
\emails{
  \upstairs{\S}These authors contributed equally.\\ \upstairs{*}Corresponding author: chloc22@clc.fitus.edu.vn
}
\vspace*{0.2in}
\end{center}

\begin{abstract}
Legal reasoning requires not only high accuracy but also the ability to justify decisions through verifiable and contestable arguments. However, existing Large Language Model (LLM) approaches, such as Chain-of-Thought (CoT) and Retrieval-Augmented Generation (RAG), often produce unstructured explanations that lack a formal mechanism for verification or user intervention. To address this limitation, we propose \textit{Adaptive Collaboration of Argumentative LLMs (ACAL)}, a neuro-symbolic framework that integrates adaptive multi-agent collaboration with an Arena-based Quantitative Bipolar Argumentation Framework (A-QBAF). ACAL dynamically deploys expert agent teams to construct arguments, employs a clash resolution mechanism to adjudicate conflicting claims, and utilizes uncertainty-aware escalation for borderline cases. Crucially, our framework supports a Human-in-the-Loop (HITL) contestability workflow, enabling users to directly audit and modify the underlying reasoning graph to influence the final judgment. Empirical evaluations on the LegalBench benchmark demonstrate that ACAL outperforms strong baselines across Gemini-2.5-Flash-Lite and Gemini-2.5-Flash architectures, effectively balancing efficient predictive performance with structured transparency and contestability. Our implementation is available at: \url{https://github.com/loc110504/ACAL}.
\end{abstract}


\section{Introduction}

Legal reasoning is the intricate process by which legal principles and rules are applied to specific facts to reach a justified outcome \cite{r17,yyy}. This professional responsibility encompasses various essential sub-tasks such as identifying pertinent issues, recalling relevant laws, and interpreting complex statutes or precedents \cite{r12,yyy}. Practitioners in this field must apply established rules to specific facts, draw logical conclusions, and present persuasive arguments to support their claims. In recent years, Large Language Models (LLMs) have demonstrated remarkable progress on these legal reasoning tasks. Researchers have shown that equipping these models with domain knowledge and specific reasoning strategies can significantly boost performance \cite{r6,r11,r12}. Recently, researchers have also begun exploring multi-agent LLM frameworks that simulate debates or judicial deliberations to tackle decision-making \cite{r3,r13}. By having multiple agents critique and refine each other's arguments, these approaches can improve factual accuracy and overall robustness \cite{r2,r25}.


However, existing approaches to legal reasoning with LLMs fall short of these requirements in different ways. Prompting-based methods, including few-shot prompting and chain-of-thought (CoT) reasoning \cite{r6,r10,r12}, often produce free-form explanations that are difficult to verify or systematically contest. Retrieval-augmented generation (RAG) improves factual grounding by injecting external legal texts \cite{r1,r13,r10}, but the final decision logic remains implicit and opaque. Multi-agent debate (MAD) frameworks introduce diversity of perspectives, yet their performance gains are inconsistent across tasks, and debate transcripts themselves do not constitute a formally contestable reasoning structure.

Furthermore, frontier models often produce invalid arguments or irrelevant citations when facing complex reasoning demands. This limitation raises concerns about trust, fairness, and accountability in high-stakes domains.
Meanwhile, regulators increasingly require transparency and contestability as reflected in global directives such as the EU AI Act \cite{act2024eu}, and Canada's Directive on Automated Decision-Making \cite{board_board_2019}.
Consequently, prior approaches either optimize predictive performance without accountability or provide explanations that lack a principled mechanism for inspection and dispute.

To address these challenges, we propose \textit{Adaptive Collaboration of Argumentative LLMs (ACAL)}, a structured, neuro-symbolic, multi-agent AI framework for explainable, contestable legal reasoning.
Our contributions are summarized as follows:
\begin{enumerate}
    \item We propose ACAL, an adaptive multi-agent legal reasoning framework that integrates argumentative LLMs (ArgLLMs) \cite{freedman2025argumentative} with adaptive role selection in \textit{arena-based (quantitative) bipolar argumentation (A-(Q)BAF)} framework, enhanced by \textit{clash resolution (CR)} and \textit{uncertainty-aware escalation (UAE)} to yield more accurate and decisive legal judgments.
    \item Our proposed ACAL supports the human-in-the-loop (HITL) contestation workflow, enabling users to directly interrogate and revise the underlying reasoning process, with changes formally propagated to influence the final outcome.
    \item We empirically demonstrate on LegalBench~\cite{guha2023legalbench} that ACAL outperforms strong prompting, RAG, CoT, and MAD baselines in predictive performance, while simultaneously providing structured transparency and contestability not supported by prior methods.
\end{enumerate}
\section{Related Work}

\subsection{LLM-based Approaches for Legal Reasoning}
The application of AI in the legal domain has transitioned from early expert systems to modern data-driven approaches leveraging LLMs. Specialized models such as ChatLaw \cite{r5} and LegalMind \cite{r8} have refined these capabilities by integrating knowledge graphs and reinforcement learning to improve factuality and process optimization. Foundational techniques like RAG are now standard for grounding LLM responses in authoritative external sources. Systems use RAG to query everything from ethical rules \cite{r1} and tax codes \cite{r10} to complex statutory frameworks and case law precedents \cite{r21}. Concurrently, CoT prompting is widely used to elicit multi-step reasoning, with some frameworks developing Law-specific CoT variants \cite{r6} or logical-semantic integration models \cite{r11} to cultivate more robust deductive pathways.
To further improve robustness, researchers have introduced multi-agent architectures in which agents assume distinct roles, such as judge, plaintiff, and defendant, to simulate adversarial legal discourse \cite{r25}. These systems, which range from simulating courtroom debates \cite{r4} to collaborative law-making \cite{r7}, leverage iterative critique to refine arguments and reach a consensus. 

However, despite these advancements, significant performance gaps remain. Recent studies highlight a ``reasoning paradox'' in which models that consume more computational resources on hierarchical tasks often degrade in performance, struggling to distinguish between surface-level facts and deeper legal distinctions \cite{r22}. Furthermore, while these advanced methods can improve accuracy over single-model baselines \cite{r3}, the debate transcripts or free-form explanations they produce lack a formal structure. This makes their reasoning difficult to systematically verify or challenge, creating a critical accountability gap that purely performance-driven architectures fail to address.

\subsection{Explainability and Contestability in LLM-based Legal Reasoning Systems}

Explainability in legal AI is fundamentally concerned with providing rational justification for stakeholders, a standard that is distinct from the technical goal of describing a model's internal mechanics. Current frameworks attempt to provide this justification by making their reasoning process transparent. For instance, systems grounded in RAG offer explainability by providing a citation mechanism that links claims to specific source documents \cite{r13,r24}. Other approaches utilize CoT techniques to generate structured, step-by-step reasoning chains that mimic judicial logic \cite{r21,r22}. More advanced systems even formalize interactions into computational argumentation graphs using frameworks like Toulmin's model to create a verifiable reasoning trail \cite{r15,r25}. However, the presence of these explanatory artifacts does not guarantee their reliability or provide a mechanism for recourse. One recent evaluation of frontier models found that over 60\% of generated judicial analyses contained invalid arguments and more than half included irrelevant citations, despite their structural coherence \cite{r21}. Consequently, there is a growing demand to move beyond passive explainability towards contestability, an interactive paradigm that empowers users to actively challenge and correct the model's reasoning process rather than merely observing it \cite{r1,r15}. 

Contestable AI (CAI) extends the principles of XAI by incorporating safeguards and providing explicit pathways for users to challenge and revise a system's conclusions \cite{xxx,www}. The integration of computational argumentation has become central to operationalizing contestability, specifically within legal AI. Frameworks such as ALEX \cite{r15} employ formal schemes, such as Toulmin's model and the ASPIC+ \cite{r15} framework, to structure interactions between opposing arguments in criminal cases. This allows users to scrutinize specific nodes in a reasoning graph. Similarly, neuro-symbolic approaches such as SOLAR \cite{r2} leverage structured ontological representations to bridge the gap between natural language and symbolic reasoning for statutory analysis. This provides verifiable pathways that can withstand adversarial probing. Other work has explored contestability through structured multi-ply dialectical schemes for case-based reasoning \cite{r16} or by deriving policy guidance from regulatory mandates to help users identify the right questions to ask \cite{r24}. However, these existing legal CAI approaches primarily focus on generating static reasoning structures for post-hoc inspection. They typically lack a quantitative mechanism to dynamically weigh the strength of conflicting arguments. Furthermore, they do not support a fully interactive workflow where human interventions, such as modifying argument weights or relations, mathematically propagate to update the final legal judgment. 

Our work bridges this gap by integrating MAD within a formal quantitative argumentation framework that is directly auditable by an HITL. This transforms static explanations into dynamic and contestable decision objects.
\section{Adaptive Collaboration of Argumentative LLMs (ACAL)}
\begin{figure}[t]
  \centering
    \includegraphics[width=\linewidth]{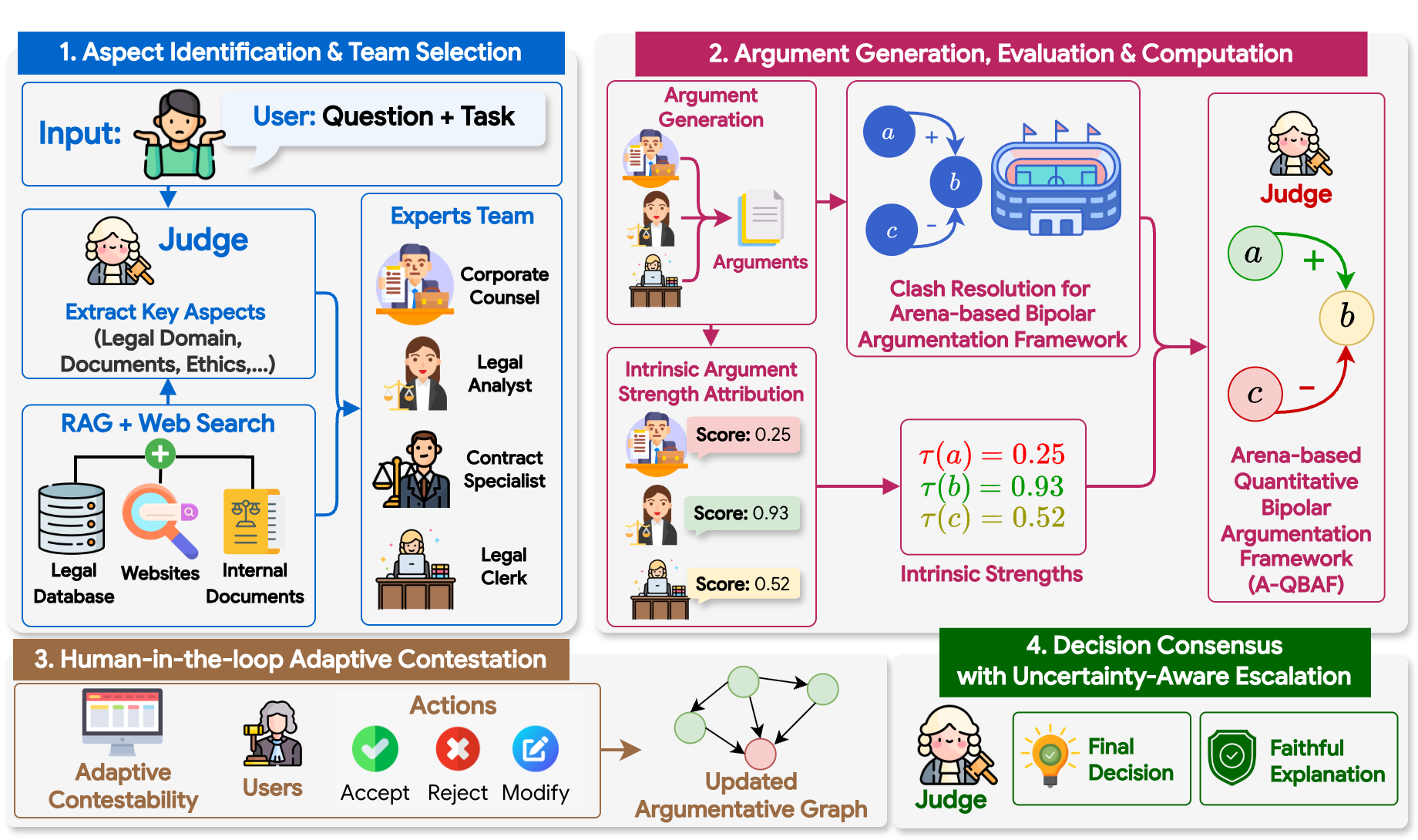}
  \caption{Our proposed Adaptive Collaboration of Argumentative LLMs (ACAL) Architecture for Legal Reasoning.}
  \label{fig:sample_fig}
\end{figure}

As shown in Figure~\ref{fig:sample_fig}, ACAL comprises four modules: adaptive expert team selection, multi-agent argument generation with arena-based clash resolution, quantitative reasoning via A-QBAF, and HITL contestation with uncertainty-aware decision consensus.
\subsection{Aspect Identification and Team Selection}
Legal reasoning faces many challenges from different domains that require specialized knowledge. A single expert might not be able to capture all perspectives in a complex case that needs different professionals to analyze it effectively. To deal with this limitation, we propose an adaptive multi-agent architecture that dynamically assembles expert teams tailored to specific legal tasks.
\subsubsection{Legal Agent Pool Definition}
We define a comprehensive pool of legal agents $\mathcal{A} = \{a_1, a_2, \ldots, a_n\}$, where each agent $a_i$ is characterized by a tuple ${r_i, E_i, P_i, S_i}$ as follows: (i) \textit{Role} $r_i$: The professional designation; 
(ii) \textit{Expertise Areas} $E_i$: A set of domain specializations;
(iii) \textit{Focus Priorities} $P_i$: Task-oriented objectives that guide reasoning;
(iv) \textit{Argument Style} $S_i$: The characteristic reasoning approach.
As shown in \autoref{tab:roles}, our framework implements 10 distinct legal roles organized into functional categories \cite{llmlawagents}.

\begin{table}[ht]
\centering
\caption{Legal roles grouped by functional category.}
\begin{tabular}{ll}
\toprule
\textbf{Category} & \textbf{Roles} \\
\midrule
Adjudication & Judge, Law Clerk / Judicial Clerk \\
Litigation \& Advocacy & Private Practice Lawyer, Prosecutor, Public Defender \\
Advisory \& Transactional & Corporate Counsel, Compliance Officer, IP Attorney \\
Research \& Support & Legal Analyst, Paralegal \\
\bottomrule
\end{tabular}
\label{tab:roles}
\end{table}

\subsubsection{Adaptive Agent Selection}
Rather than employing all agents for every case, which would be computationally expensive and potentially introduce noise when using a non-expert for a specific task, we implement an adaptive selection mechanism. Given a legal task $\mathcal{T}$ with context $c$ and claim $\phi$, the system selects two subsets:
\begin{equation}
\mathcal{A}^{+} = \text{Select}(\mathcal{A}, \mathcal{T}, c, \text{support}); \quad
\mathcal{A}^{-} = \text{Select}(\mathcal{A}, \mathcal{T}, c, \text{attack}),
\end{equation}
where $\mathcal{A}^{+}$ are agents whose expertise aligns with constructing supporting arguments for that specific case context, $\mathcal{A}^{-}$ are agents suited for generating counter-arguments. The selection function leverages an LLM to match agent expertise profiles against case characteristics.

\subsection{Argument Generation, Evaluation and Computation}
The core of our framework lies in the structured generation and computation of argument strengths using A-QBAF, which is built upon traditional QBAF \cite{qbaf}.

\subsubsection{Contextual Grounding via Hybrid RAG}
To ensure arguments are grounded in authoritative legal standards rather than generic knowledge, we integrate a hybrid retrieval module. Prior to argument generation, the system queries both a vectorized legal database for legislative texts and an external web search for recent case law. The top-$k$ most semantically relevant passages are aggregated to form the evidentiary context $c$, which conditions the subsequent reasoning of all agents to minimize hallucination and ensure accuracy.
\subsubsection{Multi-Agent Argument Generation}
Each selected agent generates arguments directly addressing the central claim $\phi$ in the context $c$. For an agent $a \in \mathcal{A}^{+} \cup \mathcal{A}^{-}$:
\begin{equation}
    \text{Args}(a) = \text{LLM}\left(a, \phi, c, \mathcal{C}, \text{type}(a)\right),
\end{equation}
where $\text{type}(a) \in \{support, attack \}$ determines whether the agent constructs evidence that the claim is $\text{True}(support)$ or $\text{False}(attack)$
Each agent typically generates 2-5 arguments, with the quantity self-determined based on available evidence, the complexity of the case.
\subsubsection{Intrinsic Strength Attribution}
Each argument $\alpha_i$ receives an intrinsic strength score $\tau(\alpha_i)$ from LLM. We designed a scoring criterion (see \autoref{tab:rubric}) that enforces strict differentiation to prevent score saturation, requiring evaluators to penalize generic statements and reward case-specific legal precision.
\begin{table}[ht]
\centering
\caption{LLM-based argument strength scoring rubric.}
\begin{tabular}{ll}
\toprule
\textbf{Score Range} & \textbf{Interpretation} \\
\midrule
$0.1 - 0.2$ & Incorrect legal analysis or misidentification of key elements \\
$0.3 - 0.4$ & Partially correct but missing critical components or overly generic \\
$0.5 - 0.6$ & Sound analysis with minor gaps or insufficient case-specific application \\
$0.7 - 0.8$ & Strong analysis with specific facts and correct legal reasoning \\
$0.9 - 1.0$ & Exceptional precision with authoritative citations and flawless logic \\
\bottomrule
\end{tabular}
\label{tab:rubric}
\end{table}

\subsubsection{LLM-based Inter-Argument Relation Identification}

Before constructing the A-QBAF graph, we must determine the pairwise relations $\mathcal{R}^{-}$ and $\mathcal{R}^{+}$ among arguments. In heuristic mode, it follows a simple rule whereby arguments sharing the same stance (both support or both attack) are treated as mutual supporters, while arguments with opposing stances are treated as mutual attackers. However, this oversimplifies legal reasoning in which two arguments on opposite sides may address entirely different legal aspects, making them logically independent rather than conflicting.

To build a better logical structure, we implemented LLM-based semantic relation identification mode. For each pair $(\alpha_i, \alpha_j)$ with $i<j$, we prompt an LLM to classify the relationship into one of three categories attack, support or neutral, along with a confidence score. To ensure robustness, any support or attack relation with confidence score $< 0.6$ is demoted to neutral. All established relations are made bidirectional, ensuring symmetric treatment in the subsequent QBAF propagation. Neutral pairs introduce no edge, keeping the graph sparse and interpretable.
For efficiency, argument pairs are analyzed in batches of $b$ pairs per LLM call (default $b = 10$), reducing the number of API calls from $\binom{n}{2}$ to $\lceil\frac{\binom{n}{2}}{b}\rceil$

\subsubsection{Clash Resolution via Arena Debating Round}
\begin{figure}[htbp]
    \centering
    \includegraphics[width=.8\linewidth]{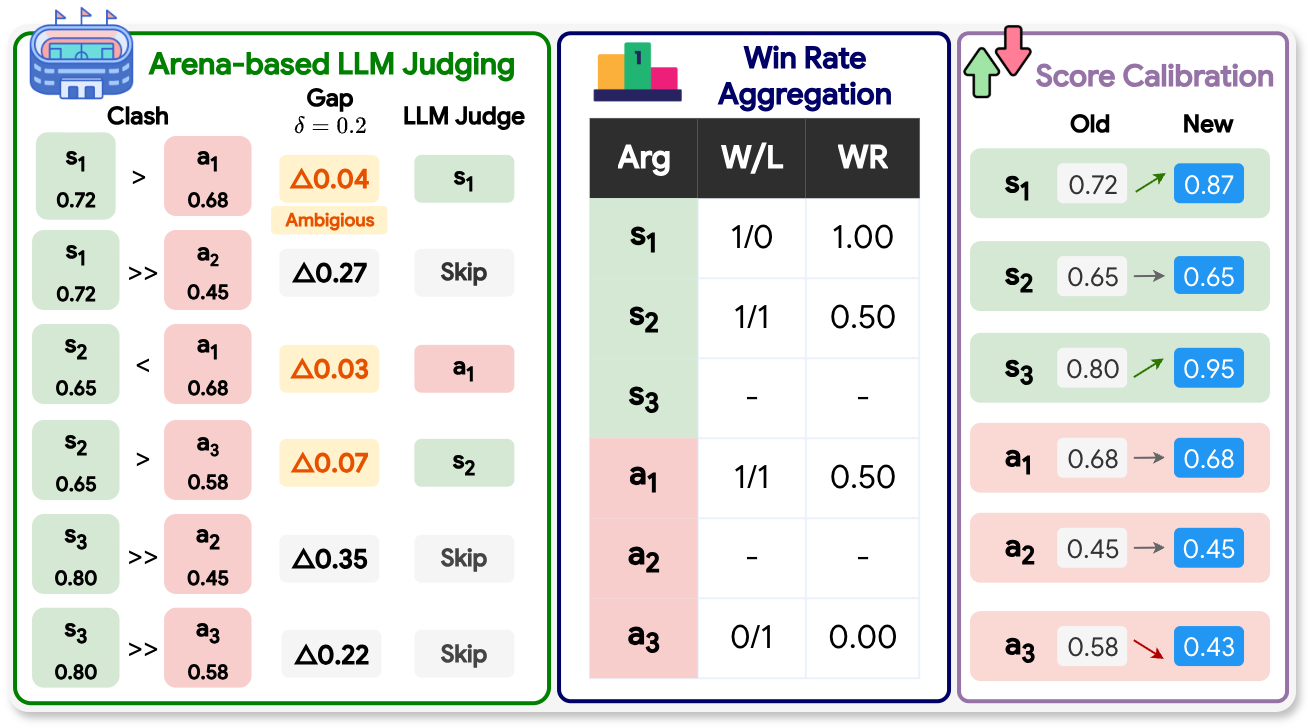}
    \caption{Illustration of an Arena Debating Round through Clash Resolution (CR).}
    \label{fig:cr}
\end{figure}

When opposing arguments have similar base scores (difference $<\delta$, default $\delta = 0.2$), we propose a \textit{clash resolution} (CR) mechanism that adjudicates between them, as illustrated in Figure~\ref{fig:cr}.
For each conflicting pair $(\alpha_s, \alpha_a)$ where $\alpha_s$ supports and $\alpha_a$ attacks the claim:
\begin{enumerate}
    \item Present both arguments to an LLM acting as a legal reasoning expert.
    \item Evaluate which argument is stronger, given case-specific facts and legal standards.
    \item Adjust scores based on win/loss outcomes across all clashes.
\end{enumerate}

The adjustment follows a symmetric formula based on the win rate $w$:
\begin{equation}
    \Delta\tau(\alpha) = \beta \cdot (2w - 1),
\end{equation}
where $\beta$ is the base adjustment magnitude and $w \in [0, 1]$ is the proportion of clashes won. This ensures winners receive bonuses while losers receive proportional penalties, maintaining score differentiation.

\subsubsection{Arena-based Quantitative Bipolar Argumentation Framework (A-QBAF)}
We formalize the argumentation structure as a QBAF $\langle \mathcal{X}, \mathcal{R}^{-}, \mathcal{R}^{+}, \tau \rangle$ where:
\begin{itemize}
    \item $\mathcal{X} = \{\phi\} \cup \{\alpha_1, \ldots, \alpha_n\}$: The set of arguments including the central claim $\phi$,
    \item $\mathcal{R}^{-} \subseteq \mathcal{X} \times \mathcal{X}$: Attack relations,
    \item $\mathcal{R}^{+} \subseteq \mathcal{X} \times \mathcal{X}$: Support relations,
    \item $\tau: \mathcal{X} \rightarrow [0, 1]$: Base strength function.
\end{itemize}

\textbf{Graph Construction}: The central claim node $\phi$ is initialized with a base score $\tau(\phi) = 0.5$ (neutral). Each argument $\alpha_i$ connects to $\phi$ via:
\begin{itemize}
    \item $(\alpha_i, \phi) \in \mathcal{R}^{+}$ if $\alpha_i$ is a support argument,
    \item $(\alpha_i, \phi) \in \mathcal{R}^{-}$ if $\alpha_i$ is an attack argument.
\end{itemize}

We compute final argument strengths using the \textit{Quadratic Energy (QE)} semantics \cite{qe}, which models score propagation as a continuous dynamical system to ensure convergence and axiomatic stability, as follows:

\textbf{Energy Calculation}: For an argument $j \in \mathcal{X}$, the total energy $E_j$ is the linear sum of the current strengths of its supporters minus its attackers
\begin{equation}
    E_{j} = \sum_{i \in \text{Sup}_j} \sigma_{i} - \sum_{i \in \text{Att}_j} \sigma_{i},
\end{equation}
where $\sigma_i$ represents the strength of argument $i$ at time $t$.

\textbf{Impact Function}: The impact of energy is governed by a quadratic function $h(x)$, which is continuously differentiable and strictly monotonic for positive values:
\begin{equation}
    h(x) = \frac{\max\{x,0\}^{2}}{1 + \max\{x,0\}^{2}},
\end{equation}

\textbf{Equilibrium Strength}: 
The final propagated strength $\sigma^*_j$ (also denoted as $\sigma(j)$ in subsequent sections) is the equilibrium state $\sigma^* = \lim_{t \to \infty} \sigma(t)$. At equilibrium, the strength satisfies:
\begin{equation}
    \sigma_{j}^{*} = \tau(j) + (1 - \tau(j)) \cdot h(E_{j}) - \tau(j) \cdot h(-E_{j}),
\end{equation}
where $\tau(j)$ is the argument's initial base weight. This ensures that attacks and supports have a symmetrical (dual) impact on the argument's strength.

\subsection{Human-in-the-loop (HITL) Contestation}
\label{sec:hitl}

While structured argumentation improves transparency, legal decision-making additionally requires \emph{contestability}: affected parties and legal professionals must be able to challenge the system’s reasoning, correct errors, and meaningfully influence the outcome.
Accordingly, we introduce a role-based HITL contestation module that turns our A-QBAF graph from a static explanation into an editable, revisable decision object.

\subsubsection{Contestation Artifacts}
Given the argumentation structure $\langle \mathcal{X}, \mathcal{R}^{-}, \mathcal{R}^{+}, \tau \rangle$ and the
computed degrees $\sigma(\cdot)$, the system generates a \emph{participation summary} and an \emph{argument dashboard}. Each argument card displays:
(i) the argument text and stance (support/attack); 
(ii) originating agent role $r_i$;
(iii) links to the specific evidence passages used (statutes, precedents, case facts);
(iv) the intrinsic score $\tau(\alpha)$ and propagated score $\sigma(\alpha)$; 
and (v) its local neighborhood in the A-QBAF (incoming supporters/attackers), plus a short ``why it matters'' trace describing its influence on $\sigma(\phi)$.

\subsubsection{Unified Contestation Workflow}
To support contestability in a single, user-facing workflow, the interface provides a guided
contestation process centered on \emph{what is being challenged} in the system’s reasoning:
\emph{(a) factual issues} (incorrect or missing facts/evidence),
\emph{(b) legal rule issues} (misapplied statute/standard or wrong test),
\emph{(c) precedent issues} (irrelevant/overruled/misread authority),
\emph{(d) missing exceptions/defenses} (e.g., consent, necessity, limitation period), and
\emph{(e) procedural fairness concerns} (lack of notice, opportunity to respond, or mitigation).
Selecting a contestation type triggers targeted prompts that either (1) request additional
supporting materials from the user or (2) instruct the relevant agents to regenerate, refine,
or rebut arguments under the user’s contestation claim. All resulting updates are reflected in
the editable argument record (arguments, relations, and scores) with an auditable change log,
ensuring contestations can materially affect the propagated claim score rather than remaining
as post-hoc feedback.

\subsubsection{Allowed Human Interventions}
During contestation, users can provide feedback on the argument record by (i) accepting or rejecting arguments; (ii) editing arguments to correct facts or refine the legal rationale; and (iii) adding missing arguments (with citations when available). Users may also suggest adjustments to an argument’s strength or its support/attack relation when the structure is clearly mis-specified.

\paragraph{\textbf{Update and recomputation}}
All accepted edits are applied to yield an updated A-QBAF $(\mathcal{X}^H,\mathcal{R}^{+H},\mathcal{R}^{-H},\tau^H)$,
after which we re-run the same propagation semantics to obtain updated scores $\sigma^H(\cdot)$ and the revised claim
score $\sigma^H(\phi)$. Thus, contestation is treated as a model input that can change the final decision rather than
a post-hoc annotation.

\paragraph{\textbf{Oversight and logging}}
For high-uncertainty or high-impact cases, the system can trigger an additional review step, and all contestation actions
are recorded in an audit log (who changed what, and how it affected $\sigma(\phi)$) to support traceability.

\subsection{Decision Consensus with Uncertainty-Aware Escalation}
The final answer is derived from the claim score:
\begin{equation}
    \text{Answer} = \begin{cases} 
\text{Yes}, & \text{if } \sigma(\phi) \geq \theta, \\ 
\text{No}, & \text{otherwise}, 
\end{cases}
\end{equation}
where $\theta$ is the decision threshold (default $\theta = 0.5$). 
However, in practice, the initial LLM-based base scoring may yield near-ties between support and attack arguments (e.g., both appear comparably plausible), causing the propagated claim score to collapse toward the neutral region. 
Thus, threshold-based decisions near the boundary risk becoming arbitrary, with outcomes driven by marginal scoring noise rather than meaningful legal distinctions.
Accordingly, prior work \cite{hasan2025survey} recommends deferring or abstaining from predictions in high-uncertainty situations, often by escalating to a stronger model or a human reviewer.

Therefore, we propose the \textit{uncertainty-aware escalation (UAE)} method for borderline cases where $\sigma(\phi) \in [0.49, 0.51]$ ($\sigma(\phi)\approx 0.5$) by bypassing the base-score-limited decision rule and invoking a \textit{Final Judge} agent. 
The Judge performs an \emph{independent} legal analysis of the case (re-evaluating evidence, legal standards, and resolving key conflicts) and outputs a \emph{binding} decision. 
This mechanism specifically addresses the same-score saturation issue between attack vs.\ support arguments and ensures a decisive outcome under high uncertainty.

\section{Experiment Setup}
\subsection{Dataset}
We evaluate our approach on \textbf{LegalBench}~\cite{guha2023legalbench}, a collaboratively built benchmark comprising a diverse suite of legal reasoning tasks designed for in-context evaluation of large language models. In this work, we focus on two classification tasks:
\begin{itemize}
    \item \textbf{Hearsay} (\texttt{hearsay}): given a short statement, predict whether it is hearsay under FRE 801(c) (\texttt{hearsay} vs.\ \texttt{not\_hearsay}).
    \item \textbf{Courts} (\texttt{learned\_hands\_courts}): given a real-world narrative (e.g., a forum-style post), predict whether it belongs to the \texttt{courts} category in the Learned Hands taxonomy (\texttt{yes} / \texttt{no}).
\end{itemize}

\subsection{Benchmark}
We conduct all experiments using two Google models: Gemini-2.5-Flash-Lite and Gemini-2.5-Flash. For the retrieval-augmented baseline, we implement a standard RAG pipeline in which documents are embedded with OpenAI \texttt{text-embedding-3-large}, extracting the top-$k$ ($k=5$) most relevant chunks to provide additional context to the generator before making a prediction. Regarding our proposed framework, we set the base adjustment magnitude $\beta = 0.15$ for the clash resolution mechanism, a value determined empirically to yield optimal performance.

We compare against commonly-used prompting and multi-agent baselines for legal and general reasoning: \textbf{SP}~\cite{sp} (few-shot standard prompting) applies a fixed instruction template with a small set of in-context demonstrations; \textbf{CoT}~\cite{cot} encourages intermediate reasoning steps before producing the final label; \textbf{RAG}~\cite{rag} augments the input with retrieved legal evidence (e.g., relevant statutory or doctrinal snippets) and conditions predictions on this external context; and \textbf{MAD}~\cite{mad} uses a multi-agent debate setting with three agents over two rounds of critique and revision to reach a final decision. 

For both tasks, we report Accuracy (Acc), Precision (Prec), Recall (Rec), and Macro-F1 (F1). Macro-F1 is computed by averaging class-wise F1 scores, ensuring equal weight across classes and making the evaluation more robust to class imbalance.

\section{Results}

\newcolumntype{C}{>{\centering\arraybackslash}X}
\begin{table}[h]
\centering
\small
\caption{Comparative results on Learned Hands Courts and Hearsay from LegalBench. All metrics are reported as percentages. The highest scores are in \textbf{bold}.}
\label{tab:main}
\renewcommand{\arraystretch}{1.3}
\setlength{\tabcolsep}{5pt}
\begin{tabularx}{\textwidth}{>{\raggedright\arraybackslash}p{1.3cm}|>{\centering\arraybackslash}p{2cm}|*{4}{C}|*{4}{C}}
\hline
\multirow{2}{*}{\textbf{Model}} & \multirow{2}{*}{\textbf{Method}} &
\multicolumn{4}{c|}{\textbf{Learned Hands Courts}} &
\multicolumn{4}{c}{\textbf{Hearsay}} \\ \cline{3-10}
 & & \textbf{Acc} & \textbf{Prec} & \textbf{Rec} & \textbf{F1} &
     \textbf{Acc} & \textbf{Prec} & \textbf{Rec} & \textbf{F1} \\ \hline
     \hline
\multirow{5}{1.3cm}{\raggedright\textbf{Gemini-2.5-Flash-Lite}}
 & SP   & 57.8 & 63.1 & 57.8 & 53.1 & 69.2 & 73.6 & 71.5 & 68.9 \\ \cline{2-10}
 & CoT  & 69.3 & 69.3 & 69.2 & 69.2 & 69.2 & 74.7 & 71.8 & 68.7 \\ \cline{2-10}
 & RAG  & 70.3 & 70.5 & 70.3 & 70.3 & 71.3 & 70.8 & 70.9 & 70.9 \\ \cline{2-10}
 & MAD  & 69.8 & 69.8 & 69.8 & 69.8 & \textbf{74.5} & 75.8 & 72.4 & 72.7 \\ \cline{2-10}
\rowcolor{gray!10} \cellcolor{white} & ACAL (Ours) & \textbf{70.8} & \textbf{71.2} & \textbf{70.8} & \textbf{70.7} &
          \textbf{74.5} & \textbf{77.1} & \textbf{76.3} & \textbf{74.4} \\ \hline
\multirow{5}{1.3cm}{\raggedright\textbf{Gemini-2.5-Flash}}
 & SP   & 65.1 & 72.5 & 65.1 & 62.0 & 75.5 & 75.7 & 74.2 & 74.5 \\ \cline{2-10}
 & CoT  & 72.3 & 74.1 & 72.6 & 71.3 & 77.2 & 77.4 & 75.6 & 75.9 \\ \cline{2-10}
 & RAG  & 75.0 & \textbf{78.1} & 75.0 & 74.3 & 75.5 & \textbf{79.5} & 72.8 & 73.0 \\ \cline{2-10}
 & MAD  & \textbf{75.5} & 76.8 & \textbf{75.5} & 75.2 & \textbf{77.6} & 78.5 & 76.1 & 76.5 \\ \cline{2-10}
\rowcolor{gray!10} \cellcolor{white} & ACAL (Ours) & \textbf{75.5} & 76.4 & \textbf{75.5} & \textbf{75.3} & 76.7 & 77.1 & \textbf{77.6} & \textbf{76.7} \\ \hline
\end{tabularx}
\end{table}

\subsection{Quantitative Analysis}
Table \ref{tab:main} presents the comparative results of our proposed framework, ACAL, against four baselines (SP, CoT, RAG, and MAD) on the \textit{Learned Hands Courts} and \textit{Hearsay} tasks from LegalBench. We evaluate performance across two backbone models: Gemini-2.5-Flash-Lite and Gemini-2.5-Flash. As shown in the table, ACAL demonstrates superior or highly competitive performance across both model architectures.

\textbf{Gemini-2.5-Flash-Lite:}
In the resource-constrained setting, ACAL achieves the best overall performance. On the \textit{Learned Hands Courts} dataset, our method surpasses all baselines, achieving the highest Accuracy (70.8\%) and F1-score (70.7\%). Similarly, on the \textit{Hearsay} dataset, ACAL outperforms the strongest baseline (MAD) in Precision (77.1\%), Recall (76.3\%), and F1-score (74.4\%), while matching the highest Accuracy (74.5\%).

\textbf{Gemini-2.5-Flash:}
Scaling to the larger model, ACAL maintains its robustness. On \textit{Learned Hands Courts}, ACAL matches the top accuracy of the Multi-Agent Debate (MAD) baseline (75.5\%) and achieves a superior F1-score (75.3\%) compared to RAG (74.3\%). Notably, on the \textit{Hearsay} dataset, ACAL achieves the highest Recall (77.6\%) and F1-score (76.7\%) among all compared methods. These results indicate that ACAL consistently outperforms standard prompting methods and remains competitive with complex retrieval and debate-based baselines.

\subsection{Explainability and Contestability Analysis}
Beyond predictive performance, ACAL addresses the opacity of traditional legal AI. Unlike baselines such as CoT or MAD, which produce unstructured text or debate transcripts, ACAL generates an A-QBAF, a structured graph where decisions are mathematically derived from explicit arguments and intrinsic scores. Furthermore, ACAL advances from passive explainability to active contestability. While RAG systems offer static citations, our HITL contestation workflow empowers users to directly audit and modify the reasoning graph. These interventions are mathematically propagated to update the final judgment, ensuring a transparent and verifiable decision-making process.

\subsection{Ablation Study}

\newcommand{\no}{\textcolor{BrickRed}{\ding{55}}}
\newcommand{\yes}{\textcolor{ForestGreen}{\ding{51}}}

\begin{figure}[t]
\centering
\captionsetup{margin=0pt}
\begin{minipage}[t]{0.48\linewidth}
\centering
\renewcommand{\arraystretch}{1.25}
\vspace{0pt}
\captionof{table}{Ablation study of our two proposed modules, Clash Resolution (CR) and Uncertainty-Aware Escalation (UAE), on Gemini-2.5-Flash-Lite (Hearsay).}
\label{tab:ablation_module}
\begin{tabular}{cc|cccc}
\hline
\textbf{CR} & \textbf{UAE} 
& \textbf{Acc} & \textbf{Prec} & \textbf{Rec} & \textbf{F1} \\
\hline
\no & \no & 64.9 & 69.8 & 67.5 & 64.4 \\
\no & \yes & 62.8 & 72.5 & 66.4 & 61.2 \\
\yes & \no & 72.8 & 75.1 & 74.5 & 72.8 \\
\rowcolor{gray!10} \yes & \yes & \textbf{74.5} & \textbf{77.1} & \textbf{76.3} & \textbf{74.4}  \\
\hline
\end{tabular}

\raggedright 
\end{minipage}%
\hfill
\begin{minipage}[t]{0.5\linewidth}
\centering
\vspace{0pt}
\captionof{table}{Ablation study of parameter $\beta$ (base adjustment magnitude) on Gemini-2.5-Flash-Lite (Learned Hands Courts).}
\label{tab:ablation-beta}
\setlength{\tabcolsep}{6pt}
\renewcommand{\arraystretch}{1.2}
\begin{tabular}{c|cccc}
\hline
\textbf{$\beta$} & \textbf{Acc} & \textbf{Prec} & \textbf{Rec} & \textbf{F1} \\
\hline
0.05 & 69.2 & 69.2 & 69.1 & 69.1 \\
0.10 & 70.0 & 70.4 & 69.9 & 69.8 \\
\rowcolor{gray!10} 0.15 & \textbf{70.8} & \textbf{71.2} & \textbf{70.8} & \textbf{70.7} \\
0.20 & 70.2 & 70.5 & 70.2 & 70.0 \\
0.25 & 69.1 & 69.6 & 68.1 & 68.0 \\
\hline
\end{tabular}
\end{minipage}
\end{figure}

\paragraph{\textbf{Clash Resolution (CR) and Uncertainty-Aware Escalation (UAE) Mechanism}} To validate the effectiveness of our proposed modules, we conducted an ablation study on the \textit{Hearsay} dataset using Gemini-2.5-Flash-Lite, effectively isolating the impact of the CR mechanism and the UAE strategy. 
This ablation study also serves as a comparative evaluation against vanilla ArgLLMs (without CR and UAE) \cite{freedman2025argumentative}, while adopting QE semantics for final argument strengths computation.

As shown in Table \ref{tab:ablation_module}, the results identify CR as the primary driver of performance, yielding a substantial $7.9\%$ increase in accuracy ($64.9\% \rightarrow 72.8\%$) when applied independently, which confirms the necessity of resolving score saturation in LLM-generated arguments. Interestingly, deploying UAE in isolation negatively impacts performance ($62.8\%$), suggesting that uncertainty estimation is unreliable without the calibration provided by CR. However, the full ACAL framework achieves the highest accuracy ($74.5\%$) and F1-score ($74.4\%$), demonstrating a complementary effect in which CR establishes the argument structure required for UAE to operate effectively in borderline cases.

\paragraph{\textbf{Base Adjustment Magnitude $\beta$}} We investigated the sensitivity of the hyperparameter $\beta$ (base adjustment magnitude), which governs the intensity of score updates during Clash Resolution. As detailed in Table~\ref{tab:ablation-beta}, performance on the \textit{Learned Hands Courts} dataset exhibits a distinct bell-shaped trend, gradually improving as $\beta$ increases from 0.05 and peaking at $\beta=0.15$ with the highest Accuracy (70.8\%) and F1-score (70.7\%). However, increasing $\beta$ beyond this threshold results in performance degradation. This finding suggests that while a moderate adjustment is essential to effectively differentiate between conflicting arguments, an excessively aggressive magnitude ($\beta \geq 0.20$) introduces volatility, leading the final propagated scores to become overly sensitive to individual clash outcomes rather than reflecting the holistic argument structure.
\section{Conclusion}
In this paper, we propose ACAL, a neuro-symbolic framework designed to address the critical need for performance and contestability in automated legal reasoning. By integrating adaptive multi-agent collaboration with an A-QBAF, ACAL successfully transforms unstructured LLM outputs into formal, verifiable reasoning graphs.
Our empirical evaluation on the LegalBench benchmark demonstrates that ACAL achieves competitive performance, surpassing robust baselines including CoT and RAG across both Gemini-2.5-Flash-Lite and Gemini-2.5-Flash models. Crucially, beyond standard performance metrics, ACAL bridges the accountability gap by enabling a HITL contestability workflow. This allows stakeholders to directly audit and intervene in the reasoning process, ensuring that legal judgments are not only accurate but also justifiable and aligned with emerging regulatory standards for high-stakes AI. Future work will focus on optimizing the computational efficiency of the multi-agent architecture to reduce inference costs without compromising reasoning depth. Additionally, we aim to validate the ACAL framework's generalizability by extending the adaptive agent pool to a broader range of complex legal tasks and other high-stakes domains that require contestable decision-making.

\printbibliography[heading=subbibintoc]

\newpage
\appendix
\section{Case Study of ACAL in Legal Reasoning}
\begin{figure}[h]
    \centering
    \includegraphics[width=\linewidth]{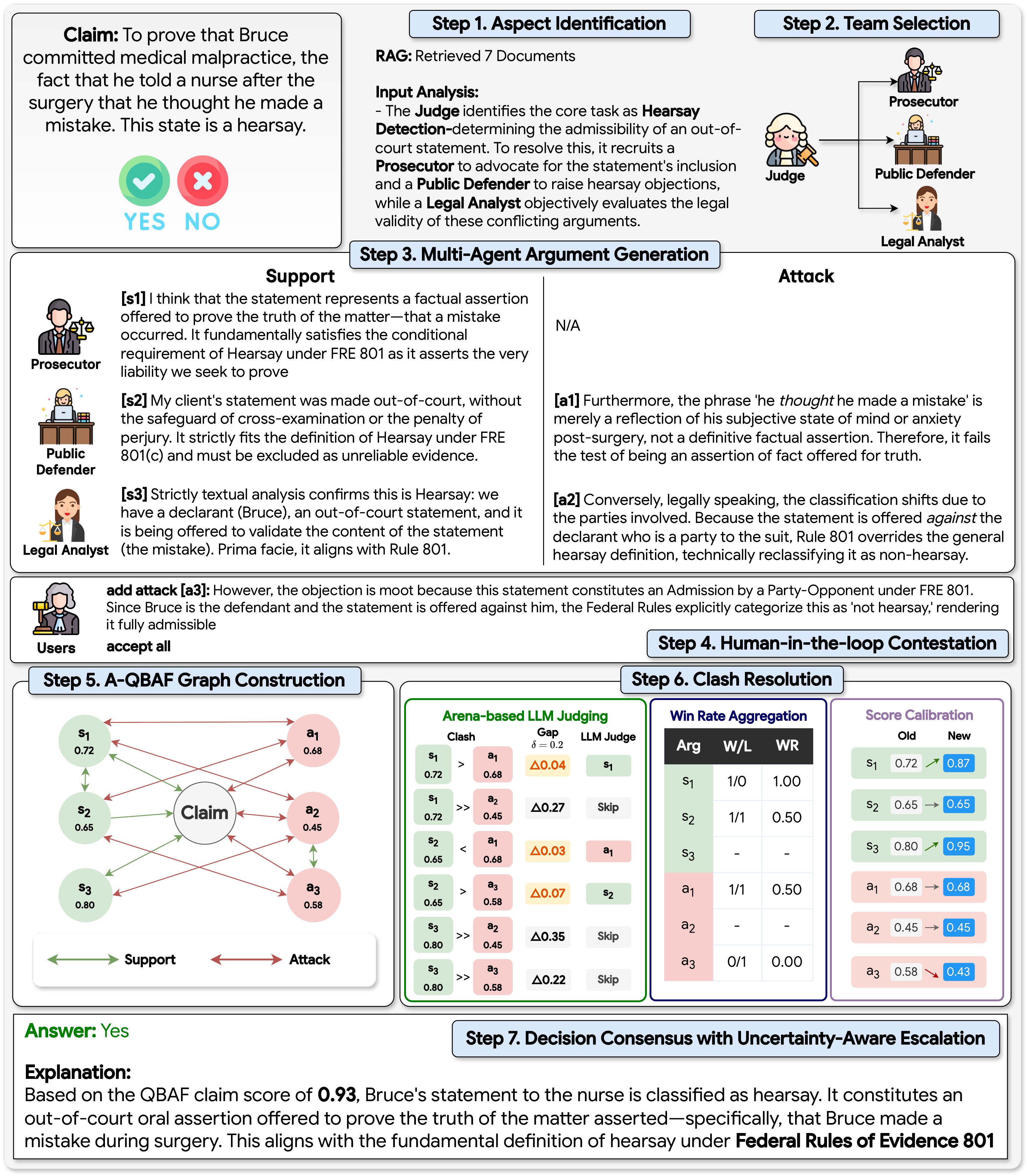}
    \caption{Illustrative Case Study of ACAL on the LegalBench Hearsay Task.} 
    \label{fig:placeholder}
\end{figure}

This appendix presents an illustrative case study of the ACAL framework applied to a complex legal reasoning scenario for a sample in the Hearsay task from LegalBench. Figure~\ref{fig:placeholder} demonstrates the end-to-end neuro-symbolic workflow, including: (1-2) Case Aspect Identification and Adaptive Expert Team recruitment; (3) Multi-Agent Argument Generation; (4) HITL contestation where a user explicitly modifies the reasoning graph; (5-6) A-QBAF Graph Construction and Clash Resolution for score calibration; and (7) Final Answer Generation with faithful explanation.
\end{document}